# Opium in science and society: Numbers

Julian N. Marewski* & Lutz Bornmann**+

*Both authors contributed equally.*

*Université de Lausanne

Faculty of Business and Economics

Quartier UNIL-Dorigny, Bâtiment Internef

1015 Lausanne, Switzerland.

E-mail: julian.marewski@unil.ch

**Division for Science and Innovation Studies

Administrative Headquarters of the Max Planck Society

Hofgartenstr. 8,

80539 Munich, Germany.

E-mail: bornmann@gv.mpg.de

+ Corresponding author

**Abstract**

In science and beyond, numbers are omnipresent when it comes to justifying different kinds of judgments. Which scientific author, hiring committee-member, or advisory board panelist has not been confronted with page-long "publication manuals", "assessment reports", "evaluation guidelines", calling for $p$-values, citation rates, $h$-indices, or other statistics in order to motivate judgments about the "quality" of findings, applicants, or institutions? Yet, many of those relying on and calling for statistics do not even seem to understand what information those numbers can actually convey, and what not. Focusing on the uninformed usage of bibliometrics as worrysome outgrowth of the increasing quantification of science and society, we place the abuse of numbers into larger historical contexts and trends. These are characterized by a technology-driven bureaucratization of science, obsessions with control and accountability, and mistrust in human intuitive judgment. The ongoing digital revolution increases those trends. We call for bringing sanity back into scientific judgment exercises. Despite all number crunching, many judgments – be it about scientific output, scientists, or research institutions – will neither be unambiguous, uncontroversial, or testable by external standards, nor can they be otherwise validated or objectified. Under uncertainty, good human judgment remains, for the better, indispensable, but it can be aided, so we conclude, by a toolbox of simple judgment tools, called heuristics. In the best position to use those heuristics are research evaluators (1) who have expertise in the to-be-evaluated area of research, (2) who have profound knowledge in bibliometrics, and (3) who are statistically literate.

> *Edinburgh, 1 Janurary 1789* Many people were at first surprised at my using the words, *Statistics* and *Statistical*… In the course of a very extensive tour, through the norther parts of Europe, which I happened to take in 1786, I found that in Germany they were engaged in a species of political inquiry to which they had given the name of *Statistics*. By statistical is meant in Germany an inquiry for the prupose of ascertaining the political strength of a country, or questions concerning matters of state; whereas the idea I annexed to the term is an inquiry into the state of a country, for the purpose of ascertaining the *quantum of happiness enjoyed by its inhabitants and the means of it future improvement*.
>
> Sir John Sinclair at the completion of his *Statistical Account of Scottland*
>
> Hacking (1990, p. 16)

# Introduction

Once, one of us spent a day in Greding, a small German village. While wandering around the medieval houses, he accidently ran into a curiosity, dating back to the 14$^{th}$ century: a large pile of bones and skulls with a sign attached to those remnants of about 2,500 humans, stating: "*What we are, you will turn into. What you are, we have been.*"

How many of us live without reflecting that they might soon die, or that their children, fathers, and sisters will turn cold and decompose – today, tomorrow, or in some other uncertain future? How many scientists think about death other than as a subject of research? In peaceful developed countries, death and dying are located in the distant, in those "poor others" who are, unfortunately, suffering from it – in wars, several thousand miles away, or in the sterile, isolated stations of hospitals around the corner. This was not always so. The uncertainty of life ikely occupied a more prominent place in collective consciousness, for example, 2,000 years ago in Roman households, and other times of high infant mortality, and recurring epidemics (e.g., Malaria in Rome; Scheidel, 1994).

It seems as if the achievements of science – including modern medicine – veil death's skeleton, and so does, most likely the technology-driven culture dominating Western societies. Yet, are people less afraid of life's uncertainty? The fear of death, so one might think, is there, but perhaps its targets have morphed. Today, doctor's dread law suits from their patients, and John Q. Public trembles at the prospect of not being able to pay back his credit. Young scientists fear to lose their jobs, and senior scientists fear the next bibliometric report. Yet, just as thousands of years ago, fear continues to be an emotion that targets something near-present or future, that is, potentially uncertain happenings. And just like in the past, also nowadays people seek relief, in one way or the other, from fear's clinging grips. In extreme cases, remedies against fear include pharmaceuticals (e.g., selective serotonin reuptake inhibitors) or psychotherapy. In less severe ones, the job is done by sports, meditation, or the products of the well-being and insurance industries (e.g., unemployment insurances). Common to most fear relievers is that they aid to shape thinking processes and belief systems.

The focus of this essay is on such powerful mean to shape – and often cloud – people's judgments: *numbers*. Numbers have been turned into, so we argue, in allusion to Karl Marx' famous statement about religion, the new opium of the people, namely when it comes to making judgments about uncertain things. Numbers affect the general public as much as those experts who dedicate their lives to produce, process, and scrutinize numbers: scientists. Give people numbers, and they will have something to hold on to, to be blinded by, or to argue against. Numbers influence what we fear, and also how we reason about our fears – be it when trying to assess the probability of dying in an airplane crash ("Just 1 in several million!"), our life expectancy at age 50 ("Still about 30 years of life remain!"), or when calculating how much unemployment benefits will enter our bank accounts if we don't happen to get tenure ("80% of the last salary for 1 year!"). Our central theses are as follows:

1. Numbers dramatically change how we make judgments under uncertainty – not only about fear-inspiring events, but also about multiple other aspects of science and society. Particularly, during the past century, $p$-value statistics and other numbers have transformed the way social scientists evaluate theories and findings (e.g., Gigerenzer et al., 1989). And currently, bibliometric statistics transform judgments about scientists, their work, and scientific institutions themselves – namely when it comes to "assessments" of "productivity", "value", or "quality": *quantitative science evaluation*. The ongoing digitalization of Western societies potentiates this trend.

2. In science and beyond, the routine uses of numbers cement social conventions (e.g., citation rates, *h*-indices, and *p*-values to evaluate scientists and their findings), aid governance and fuel beraucracization (e.g., governments and research instutions establishing databases to measure scientific "output"; Kostoff, 1998). Numeric indicators seemingly aid establishing "objective" facts and "certainties". Those, in turn, serve to justify decisions (e.g., about funding scientific work or hiring senior scientists), and, if need to be, also put decision makers (e.g., scientists, administrators, politicians) in a position to defend themselves (e.g., against being accused to have made arbitrary, biased, nepotistic, or otherwise flawed funding decisions). The ongoing digitalization shapes this trend, too.

3. The advance of numbers as substitute for judgment is propelled by rather old ideals, including those of "rational, analytic decision making", dating back, at least, to the Enlightenment. The past century has seen a new twist of those ideals, with much work in the decision sciences documenting how human judgment deviates from alleged goldstandards for rationality, this way trying to establish intuition's flawed nature (e.g., Kahnemann, Slovic, & Tversky, 1982). While the accent on human irrationality has further fuelled the advance of "objectifiers" – including measurement and numbers – it has also brought harm, namely by likely contributing to the blind – and often even mindless – routine use of seemingly "objective" indicators.

4. We point towards antidotes against the harmful sideeffects of the increasing quantification of science evaluation: while the mindless use of indicators for seemingly "objective" evaluation is nowadays prevalent in virutally all branches of science, within the decision sciences themselves, a novel view of how humans ought to make good judgments under uncertainty has gained impetus (Gigerenzer & Gaissmaier, 2011; Gigerenzer, Todd, & ABC Research Group, 1999). In line with that view, we point out how the mindless use of numbers can be overcome if science evaluators and consumers of science evaluations learn to rely on a repertoire of simple decision strategies, called *heuristics*. Being statistically literate aids all three: (i) to know when to use which heuristic, (i) to understand when good human judgment ought to be trusted even when numbers speak against that judgment, and (iii) to realize why good human judgment and intuition is, at the end of the day, what matters, and that even when there is no number attached to it. Just like most of us immideatly grasp what the dark, empty eye sockets of the skulls of Greding and their message "*What we are, you will turn into. What you are, we have been*" tell us, all without any number attached to those remnants of fathers, mothers, and children.

In what follows, we will first sketch out the historical contexts and societal trends that have come with the increasing quantification of science and society. Second, we will turn to those developments' latest outgrowth: the use of bibliometric statistics for research evaluation purposes. We close by calling for bringing sanity back into scientific judgment exercises and explain how relying on a repertoire of heuristics might aid doing so.

# Significant Numbers

**Numbers aid governance and fuel bureaucratization**

The quest for numbers is not new. Numbers, written on papyrus, coins or mile stones aid to govern societies and their activities – ranging from trade to war – since thousands of years. The Roman empire offers examples (Bowman, 2013; Vindolanda Tablets Online, 2018). France, with Napoleon's appetite for numbers (Bourguet, 1987), and the Prussian bureaucracy offer others. Numbers prepared the ground for modern welfare states, enabling to compute contributions to pension funds or sickness benefits. Numbers lent arguments to political reformers and activists, including Karl Marx (Gigerenzer, 2008; Hacking, 1990). And modern states are unthinkable without them, with today's tax offices, the *FDA*, the *CIA*, the *EPA*, and thousands of other institutions simply continuing the same old business: produce and interpret numbers to inform, make, and justify judgments. Indeed, the word statistics likely originates in states' quest for economic, demographic or other data (Gigerenzer & Marewski, 2015). "Big data" is one of the latest outgrowths of those developments, in the future possibly leading to E-governance and digital democracy or, in the worst case, to the dictatorship of digitalized numbers as foreshadowed by China's citizens score (e.g., Helbing, 2015a; Helbing, 2015b; Helbing et al., 2017; Helbing & Pournaras, 2015). A development that can be summarized by Galileo Galilei's famous lemma: measure what is measureable, and make measurable what is not; or, reformulated in political terms, control what is controllable and make controllable what is not.

**Numbers offer seemingly universal and automatic means to ends**

*Measure what is measurable* – indeed, leaving the nightmare of a government-steered or profit-driven big (data) brother aside, wouldn't it be wonderful if it were possible to measure, analyze, and monitor everything, ranging up to the costs and benefits of tax payers' investment in health care plans, environmental pollution, risks of genetic manipulation, and other thorny issues? Actually, is this not what we expect both scientists and public administrators to do? Nowadays, quite some people do.

Using analysis and reason to understand and rule the world are old ideals. They can be found in the Enlightenment, espoused by thinkers such as Immanuel Kant, Benjamin Franklin or by the enlightened Despot, Frederick the Great, King of Prussia. Gottfried Wilhelm Leibniz (1951) dream of a universal calculus that would allow translating ideas into symbols, and in so doing, permit to settle any kind of scientific dispute, beautifully illustrates those ideals, too: *instead of guessing and arguing, let us simply sit down, and calculate!* (Gigerenzer, 2014).

Modern echoes of those old ideals are *universalism* and *automatism* (Gigerenzer & Marewski, 2015). Automatic refers to establishing "neutral", "objective" measurement and quantitative evaluation procedures that – independent from the people using them – produce unbiased judgments for improving decision making. Universal refers to the dream of those automatic procedures to be applicable to all problems, independent of context. Omnipresent in all universal and automatic proceedures are numbers – numbers can be conveniently used *independent of context* (one can enumerate anything) and moreover, they seem to lend "objectivity" to observations that is *independent of the observers* (it does not matter who counts the number of words of this essay, everybody should arrive at the same number).

In scientific research, a prominent example of universalism and automatism is the routine use of null hypothesis significance testing (*NHST*) for *all* statistical inferences (Gigerenzer, 2004; Gigerenzer, Krauss, & Vitouch, 2004). Statistical inferences are judgments under uncertainty. In making those judgments, researchers in psychology, business, medicine, and economics blindly report $p < .05$, as if the $p$-value would not depend on their own intentions (see e.g., Kruschke, 2010, for dependencies of the p-value), or as if that number were equally informative for all judgment problems (see e.g., Forster, 2000; Gelman & Hennig, 2017; Gigerenzer & Marewski, 2015; Marewski & Olsson, 2009, for a discussion of when different approaches to scientific inference can be informative). For instance, an estimated 99 $p$-values were computed, on average, *per article* in the *Academy of Management Journal* in 2012 (Gigerenzer & Marewski, 2015). Similarly, *Nature*, *Science*, and other flagship journals fill their pages with $p$-values and double or triple stars (McCloskey & Ziliak, 1996; Schneider, 2015; Tressoldi, Giofre, Sella, & Cumming, 2013; Ziliak & McCloskey, 2004).

Even the devine has been assessed with numbers – actually by means of the first known test of a null hypothesis (Gigerenzer et al., 2004). In 1710, John Arbuthnott noted that "the external Accidents to which are Males subject (who must seek their Food with danger) do make a great havock of them, and that this loss exceeds far that of the other Sex" (Arbuthnott, 1710, p. 188). To counteract this trend, so he proposed "provident Nature, by the disposal of its wife Creator, brings forth more Males than Females" (p. 188), so that "every Male may have its Female" (p. 186). Using 82 years of birth records in London, John Arbuthnott asked whether chance or "Divine Providence" (p. 186) led to more male births than female ones. He concluded "that it is Art, not Chance that governs" (p. 189). The calculated "Expectation" (p. 188) of the observed excess of males, given chance was simply too small: $p = (½)^{82}$. Put in modern terms, the probability $p(D|H)$ of the observed data, $D$, given the null hypothesis, $H$, was $p > 000001$. "From hence it follows", he added, "Polygamy is contrary to the Law of Nature and Justice" (Arbuthnott, 1710, p. 189).

About 200 years later, the Duke parapsychologist Rhine used statistics to establish extra-sensory perception, *ESP*, with "statistical luminaries" of the time defending the statistics used to make the case for ESP (Gigerenzer et al., 1989, p. 243). In the same century, the renouced statistician Lindley (1983), one of the most-outspoken advocates of a universal and automatic use of numbers, namely Bayesian statistics, writes: "We are uncertain about the inflation rate next year, the world's oil reserves, or the possibility of nuclear accidents. All these can be handled by subjective probability" (pp. 10-11). And those "[b]ayesian methods are even more automatic" than Fisherian ones (Lindley, 1986, p. 6): "The Bayesian paradigm provides rules of procedure to be followed. I like to think of it as providing a recipe: a set of rules for attaining the final product. The recipe goes like this. What is uncertain and of interest to you? Call it θ. What do you know? Call it D, specific to the problem, and H, general. Then calculate p(θ|D, H). How? Using the rules of probability, nothing more, nothing less" (Lindley, 1983, p. 2). Seemingly, there are no limits to what scientists can do with numbers, and those numbers aid letting even bold claims look "objective", rational, and hence, defendable.

But not only scientists use automatic, universal statistics to make judgments; science itself is increasingly submitted to context-blind, number-driven inference. Here, an important example of universalism is the routine reliance on citation rates, *h*-indices, and impact factors for making judgments about the "productivity" or "quality" of scientists, institutions, and scientific outlets, independent of context. Context can be the discipline, the research paradigm within a discipline, or the unit investigated (be it a teaching- or research-oriented professor or institution). Automatism in research evaluation takes the disguise of legal procedures that come, for instance, with faculty evaluation exercises. Citation rates, numbers of publication, journal impact factors and other numbers should "objectively" tell, independently of who conducts the evaluation, whether a scientist is worth hiring or worth a promotion. Which hiring committee-member, grant-application reviewer or advisory board panelist has not been confronted with page-long "evaluation guidelines" calling for those numbers in order to justify inferences about job candidates and potential grantees?

What is more, thanks to the ongoing digitalization, today, automatic and universal statistics can be found everywhere. Outside of the world of research, incarnations of the old ideals of universalism and automatism are attempts to "measure" the environmental impact of different consumer goods (e.g., http://www.ghgprotocol.org/), to "assess" quality in health care (e.g., hospital ratings; see Wachter, 2016), or to "evaluate" education systems across countries (e.g. as per the *Programme for International Student Assessment, PISA*). We can learn, online, that the odds of dying from an asteroid impact are just 1 in 74,817,414 (Farber, 2013); there is an *iPhone App* that passengers can consult, prior to embarking on their flight, to check their probability of "going down". In this (brave) new digital world, is there any judgment that cannot be made from numbers?

**How numbers seem to replace judgement**

Numbers communicate reassuring objectivity, and aid to establish "facts", even when it comes to the supernatural, the divine, or the uncertain future. Indeed, nowadays, scientists, politicians, doctors, and businessmen evoke numbers rather than hunches or gut feel to motivate and justify their thoughts about scientific ideas, spending policies, diseases, or mergers. We collect data on past financial returns to defend future investments, we use past survival rates to forecast a cancer patient's chances of being alive in the future, and a scientists' publication output and citation rates serve to infer that scholar's future performance, motivating tenure decisions. We even use the absence of numbers to make judgments, such when pointing out that few are known to have died from smoking (a few decades ago), or from living close to nuclear power plants (still today). Numbers, cast into statistics, equations, and algorithms, seem to allow us to control the uncertainties of the future and to justify present-day decisions. Numbers have, for the better or the worse, replaced mere "hunches" (Hoffrage & Marewski, 2015).

This was not always so, and not in all contexts – and one does not have to go to times and places of shamanic rituals, religious fanaticism, or so-called superstitions for cases in point. Medicine, for instance, was based throughout centuries on the "medical tact" emphasizing the individual patient, with physicians objecting vehemently against summary statistics such as means. Only in the 20th century did probabilities and other numbers obtained from randomized trials displace physician's clinical judgments (Gigerenzer et al., 1989, chapter 2). Likewise, in the past century, patients' common-sense intuitions such as *Go to the doctor when it hurts!* increasingly became overwritten by numbers, with the numbers sometimes contradicting the intuitions. Relative risks, as used in statements such as *The risk of dying from breast cancer is reduced by 25% among women who participate in mammography screening!* are a case in point; conducing healthy people to participate in massive cancer screening programs (Gigerenzer, 2014; Gigerenzer, Gaissmaier, Kurz-Milcke, Schwartz, & Woloshin, 2008).

Intelligence, measured by IQ and other numbers, is an invention of the 20$^{th}$ century, too; serving, for instance, U.S. military recruitment, immigration control, and dubious policies, with deficient mental test scores offering legal grounds for limiting access to education, and in some American states, even for sterilization (Gigerenzer et al., 1989, chapter 7). To be blamed are psychologists and the law-makers who believed them, turning mental numbers into seemingly objective, easy, and universal criteria to single out "bad pennies". Multiple choice test scores, including the notorious SAT, GMAT, and other universal, automatic indicators of scholastic aptitude, were absent from university education prior to those times.

To give a final example, contrary to the Enlightenment's focus on analysis, reason, and rationality, the great thinkers of the German epoch *Sturm und Drang* (ca. 1765-1785) placed emphasis on feelings, emotions, and their outward expression. Johann Wolfgang Goethe's suicidal Werther reflects this mindset (Hoffrage & Marewski, 2015). Today, Werther would be placed into a psychiatric clinic, and self-respecting behavioral scientists prefer to *study* shamanism, religion, or the "emotional brain" rather than to letting themselves be guided by them. Not even those who study intuitive judgments – decision scientists – admit using intuition when it comes to assessing their own data and theories on intuition – and even less so when it comes to grading their students; without objective test scores, law suits loom on the horizon.

Seemingly irrational behaviors and subjective, biased cognitions, ranging from feelings to intuitions, pose, however, prominent targets for investigation when it comes to documenting, correcting, and even exploiting them. This occurs in both science and in governance, with one informing and fueling the other. Starting in the 1970's, Kahneman and Tversky's *heuristics-and-biases research program* (e.g., Kahnemann et al., 1982) brought irrational, error-prone, and faulty judgment and decision making into thousands of journal pages, and onto administrators' agendas. Many of those judgment biases were defined based on numbers; including by experiments showing how people's judgments, seemingly, violated benchmarks for rationality such as the laws of probability in Bayes theorem (Gigerenzer, Hell, & Blank, 1988; Gigerenzer & Marewski, 2015; Gigerenzer & Murray, 1987). In other research programs, statistics from behavioral studies fueled similar conclusions; namely that irrational citizens need outside help and steering. The *libertarian paternalist movement*'s emphasis on nudging ignorant people (Thaler & Sunstein, 2008) is one prominent example. Also certain notions of egoistic homo economicus – who in the absence of punishment and control will inevitably maximize his utility, potentially at cost to others – fits the widespread view that people's subjective judgments cannot be trusted.[1] The most recent outgrowth of the mistrust in good human judgement is the view that artificial intelligene will soon outperform human intelligence.

In short, the old ideals of objectivity and universality have, over centuries, come to shape numerous domains in our society, ranging from religious believes to governance and research. Those ideal's lifeblood are numbers: numbers can be applied universally and automatically, seemingly independent of context and people. The historic spread of numbers has, in more recent years, been associated with a negative conception of man's ability to make good judgements all by himself. Judgements have to be unbiased; numbers can aid to objectify and, hence, justify them.

## Science evaluation

**How numbers fuel the quest for objective, unbiased, and justifiable judgment**

In 1987, Gigerenzer published a beautiful piece on the "fight against subjectivity" (Gigerenzer, 1987, p. 11), illustrating how, historically, the use of statistics became institutionalized in psychology, namely when experimenters started to invoke them to make their claims independent (i) from themselves as well as (ii) from the human subjects they studied. The fight against subjectivity is neither unique to psychology, nor has the struggle for objectivity ended in the social sciences. Across disciplines, scientists currently use numbers to make their claims about the value of their own and other's scientific work appear independent from them and from context (see e.g., Gelman & Hennig, 2017). Academia, so we argue, is in the process of undergoing a dramatic transformation: as much as numbers have contributed to transform other aspects of science and society, they shape science evaluation.

---

[1] Utility-maximizing economic models offer an example of how the ideal of universalism is actually reflected by scientific theories themselves. In corresponding economic models, utilities are expressed numerically such that, in principle, everything (ranging from costs and benefits coming with crime and punishment, marriage, fertility, altruism, or discrimination) can be modelled with them (see e.g., Becker, 1976). And related models are not only prevealent in economics but also in other behavioral sciences, such as psychology (e.g., for learning; Anderson, 2007) or biology (e.g., for foraging; Stephens & Krebs, 1986).

What is being evaluated varies, ranging from individual journal articles to different "producers" of science, including scientists or competing departments and universities. For instance, many use, nowadays, the amount of citations a paper enlists on *Google Scholar* to find out whether that paper is worth citing and reading. Likewise, when it comes to justifying promotions of assistant professors or to allocating limited amounts of funds to competing departments, what frequently counts is the number of papers published in journals on the *Financial Times List*, the *HEC Paris List*, or in the top quarter of the *Clarivate Analytics Journal Citation Reports*.

Similar statistics serve the general public and policy makers too, with tax payer's investments in academic institutions, personnel, and research calling for indicators of success as justification. A frightening example is the *Research Excellence Framework* of the United Kingdom – a "system for assessing the quality of research in UK higher education institutions" (Research Excellence Framework, 2018), informing the general public about the quality of British science (e.g., for 2014, "the overall quality of submissions was judged 30% world-leading…, 46% internationally excellent…", Research Excellence Framework, 2014b). The public website allows to search in a database, through different institution's rating scores (e.g., "world-leading"), all the way down to individual scientists or citation counts for individual papers. From the mission statement:

- "The … higher education funding bodies will use the assessment outcomes to inform the selective allocation of their grant for research to the institutions which they fund…
- The assessment provides accountability for public investment in research and produces evidence of the benefits of this investment.
- The assessment outcomes provide benchmarking information and establish reputational yardsticks, for use within the higher education (HE) sector and for public information" (Research Excellence Framework, 2014a).

Not only in the United Kingdom, but also in many other countries (e.g. Australia or the Netherlands), do those developments at the level of society feed and are fed by smart businessmen. Companies launch an ever increasing stream of off-the-shelves, user-friendly number producers, ranging from automatic citation counts (e.g., *Google Scholar*) and network apps (e.g., *ResearchGate*) to bibliometric products (*InCites* from *Clarivate Analytics* or *SciVal* from *Elsevier*) and plagiarism detection software. Also lawyers and journalists have their share, with the public outcries *Corruption! Nepotism!* or the latent threat of court trials (e.g., from job candidates) incentivising academic institutions to implement (e.g., resource allocation) procedures that are not, first and foremost, sensible, but that are defendable. The rationale of the number-based defenses is: numbers are harder to argue with than "subjective judgments".

The quantification of science evaluation has numerous consequences. Scientists seem to get more obsessed with achieving $X$ journal publications per year than with simply producing research for the sake of producing compelling research, that is, on its *own* right (Paulus, Rademacher, Schäfer, Müller-Pinzler, & Krach, 2015). Instead of making one's line of reasoning comprehensible to other researchers, papers are cited to boost one's $h$-index, without much reflection about what those citations actually stand for (Tahamtan & Bornmann, 2018). And plagiarism software and its similarity indices invite graduate students to worry more about getting alarms from the software rather than about learning how to cite adequately. Indicators produce unintended consequences and replace thinking and insight (Waltman & van Eck, 2016).

Those development strikingly resemble what Gigerenzer and others observed, decades earlier, for the quantification of social science research itself: there, $p$-value statistics started to spread when educational and parapsychologists wanted to make their claims (e.g., about extrasensory perception) look objective, and when textbook writers (e.g., Guilford, 1942) started to sell a mishmash of Ronald A. Fisher's, Jerzy Neyman's, and Egon Pearson's competing statistical frameworks to social scientists without that those textbook writers actually understood the statistics they sold (Gigerenzer, 2004; Gigerenzer et al., 2004; Gigerenzer & Marewski, 2015). Journal editors picked up those writings. They found in *p* =.05 a seemingly universal and "objective criterion" for judging findings; and a few decades later, technology including off-the-shelf software such as *SPSS*, made assessing that criterion easy. The aftermath saw scientists worrying more about the magic .05 than about the soundness of theories, the quality of data, or the information conveyed by the *p*-value itself. In the social sciences, this way, NHST was born and institutionalized – a seemingly universal and automatic judgment procedure, applicable to all statistical problems independent of context and people. Few researchers realized or cared that they did not really understand the numbers produced by the testing, and even fewer noticed that the testing itself lacked coherent grounding in statistical theory or that there are, actually, three notions of level of significance (i.e., two from Fisher, one from Neyman and Pearson) all of which differ dramatically from NHST (Gigerenzer, 2004; Gigerenzer et al., 2004).

**How numbers fuel the bureaucratisation of science**

The quantification of science evaluation has not ended with mere efforts towards making judgments look unbiased, objective, and justifiable. The bureaucratization of science is another (equally worrisome) outcome. Modern science is frequently called *post-academic*. According to Ziman (2000), bureaucratization is an appropriate term which describes most of the processes in post-academic science: "It is being hobbled by regulations about laboratory safety and informed consent, engulfed in a sea of project proposals, financial returns and interim reports, monitored by fraud and misconduct, packaged and repackaged into performance indicators, restructured and downsized by management consultants, and generally treated as if it were just another self-seeking professional group" (p. 79). For example, principal investigators in *Advanced Grants* by the *European Research Council* (*ERC*) should be able to provide evidence for the calculation of their time spent on the ERC project – time sheets enter academia (European Research Council, 2012).

Words such as management, performance, contract, regulation, accountability, and employment had previously no place in scientific life. The vocabulary was not developed within science, but was transferred by the "modern", bureaucratized society (Ziman, 2000). The bureaucratic virus spreads through the scientific publication process itself: nowadays, many journals require hosts of (e.g., web) forms to be signed (or ticked), ranging from "conflict of interest statements", to "ethical regulations", assurances that the data to be published is "new and original", to copyright transfer agreements. Certain publication guidelines read like instructions one would otherwise find in tax forms in public administration.[2]

Research in post-academic science is characterized by less freedom. Projects are framed by proposals, employment and supervision of project staff (PhD students, post-doctoral researchers), and performance measurements which carry the risk that research is being condemned to operate in terms of "normal science" (Kuhn, 1962). The pressure to operate in terms of "normal science" leaves less space for serendipity and scientific revolutions. Explorative studies—which can lead to scientific revolutions—can raise questions that are new and not rooted in the field-specific literature; such studies can, moreover, come with unconventional approaches, and lead to unforeseeable expenditures of time (Holton, Chang, & Jurkowitz, 1996; Steinle, 2008). There is the risk that these elements are negatively assessed in grant funding and research evaluation processes, because they do not fit into "efficient" project management schemes.

In short, the four most important characteristics of post-academic science evaluation can be summarized as follows (Moed & Halevi, 2015).

**(1) Performance-based institutional funding.** In many European countries, the number of enrolled students is decreasingly and performance criteria are increasingly relevant for the amount of research funds for universities. The performance criteria are used for accountability purposes (Thonon et al., 2015). "In the current economical atmosphere where budgets are strained and funding is difficult to secure, ongoing, diverse and wholesome assessment is of immense importance for the progression of scientific and research programs and institutions" (Moed & Halevi, 2015, p. 1988).

---

[2] Ironically, several decades ago, the *Publication Manual of the American Psychological Association* indeed instructed its readers, to "treat the result section like an income tax return. Take what's coming to you, but no more" (American Psychological Association, 1974, p. 19).

**(2) International university rankings**. Universities are confronted with the results of international rankings. Although heavily criticized (Hazelkorn, 2011), politicians are influenced by ranking numbers in their strategies for funding national science systems. There are even universities (e.g. in Saudi Arabia or China) incentivising behavior to influence their positions in rankings, for instance, by awarding nominal "visiting faculty" status to highly cited researchers from universities in other countries (Bornmann & Bauer, 2015). *Clarivate Analytics* publishes annually a list of researchers (at https://clarivate.com/hcr/) who have authored the most papers in their disciplines belonging to the 1% most frequently cited papers. This list is constitutive of one of the best known international rankings, the *Academic Ranking of World Universities* (*ARWU*), also known as *Shanghai Ranking*. *The Economist* and the *Financial Times* undertake rankings of business schools and their Master, MBA, and Executive MBA programs.

**(3) Internal research assessment systems**. More and more universities and funding agencies install research information systems to collect relevant data on research "output" (Biesenbender & Hornbostel, 2016). These numbers are used to monitor performance and efficiency continually. Problems emerge if those monitoring systems change researchers' goals in an unintended way – for instance, leading them to dissect ideas into small publishable journal units, instead of creating comprehensive theory that would fill books. Equally problematic, the systems incentivize surfing the wave of "currently hot topics", well-established research paradigms and methodologies, instead of thinking out of the box, pulling together different areas of expertise, and risking for their work to be (initially) neither published nor popular or well-understood. The 2013 Nobel Prize laureate in physics, Peter W. Higgs, bluntly remarked, "Today I wouldn't get an academic job … I don't think I would be regarded as productive enough" (Aitkenhead, 2013). Indeed, prior to receiving the prize, so Higgs noted, he had become "an embarrassment to the department when they did research assessment exercises" (Aitkenhead, 2013). When requested his recent publications, "Higgs said, I would send back a statement: 'None'" (Aitkenhead, 2013).

**(4) Performance-based salaries.** In several countries (e.g. China, India, Portugal, Czech Republic) salaries are sometimes connected to publishing *X* articles in reputable journals (e.g. *Science*, *Nature*) (Reich, 2013). In business schools, it is common to link a reduced teaching load, promotions and other benefits to publishing in journals on the *Financial Times List* – the same newspaper that publishes the MBA and other rankings important to the schools' prestige and, ultimately, to profitability of their programs. Such practices widely open the doors to scientific misconduct (Bornmann, 2013b). One is not really be surprised to read that papers are bought from online brokers or that scientists pay for authorships (Hvistendahl, 2013).

**Why numbers alone are not sufficient when it comes to research evaluation**

According to Wilsdon et al. (2015), today three broad approaches are mostly used to assess research in post-academic science: the *metrics-based model*, which relies on quantitative measures (e.g. counts of publications, prices, or funded projects), peer review (e.g., of journal or grants), and the combination of both approaches. Standards in science, so the rationale of the peer review process goes, can only be established if research designs and results are assessed by peers from the same or a related field. In the past decades, the quest for comparative and continuous evaluations on a higher aggregation level (e.g., institutions, countries) has fueled preferences for the metrics-based model. Those preferences are also triggered by the overload of the peer review system: the demand for the participation in peer review processes exceeds the supply.

In the metrics-based model of research evaluation, bibliometrics has a prominent position (Schatz, 2014; Vinkler, 2010). According to Wildgaard, Schneider, and Larsen (2014) "a researcher's reputational status or 'symbolic capital' is to a large extent derived from his or her 'publication performance'" (p. 126). Bibliometric information is available in large databases (e.g. *Web of Science*, *WoS*, provided by *Clarivate Analytics* and *Scopus* from *Elsevier*) and can be used, seemingly flexibly, in many disciplines and on different aggregation levels (e.g. single papers, researchers, research groups, institutions, countries). Whereas the number of publications is used as an indicator for production (i.e., productivity), the number of citations is relied upon as proxy for quality.

However, the metrics-based model has several pitfalls (Hicks, Wouters, Waltman, de Rijcke, & Rafols, 2015). Five of those problems stem from the ways in which numbers are used (Bornmann, 2017).

**(1) Skew in bibliometric data.** Bibliometric data tend to be right-skewed, with there being only a few highly-cited publications and many publications with only a few or zero citations (Seglen, 1992). The "bibliometric law" says that there is a tendency for items (here: citations) to concentrate on a relatively small stratum of sources (here: publications; de Bellis, 2009). Citations are over-dispersed count data (Ajiferuke & Famoye, 2015). Hence, simple arithmetic means – as they are built into mean citation rates or journal impact factors (*JIFs*, measuring the mean citation impact of a journal) – should be avoided as measure of central tendency (Glänzel & Moed, 2013).

**(2) Variability in bibliometric data.** In line with the ideals of universalism and automatism, the results of bibliometric studies are typically published as if they were independent of context or otherwise invariant (Waltman & van Eck, 2016). Single highly-cited publications ("outliers" of sorts) can have a great influence on the results, and the results of bibliometric studies can vary between different samples (e.g. from different publication periods or literature databases).

**(3) Time- and field-dependencies of bibliometric data.** Many bibliometric studies are based on bare citation counts, although these numbers cannot be used for cross-field and cross-time comparisons (of researchers or universities). Different publication and citation cultures lead to different average citation rates in the fields – independently of the quality of the publications.

**(4) Language effect in bibliometric data.** In bibliometric databases, English publications dominate. Since English is the most frequently used language in science communication, the prevalence of English publications comes as no real surprise. However, the prevalence can influence research evaluation in practice. For example, there is a language effect in citation-based measurements of university rankings, which discriminates, particularly, against German and French institutions (van Raan, van Leeuwen, & Visser, 2011). Publications not written in English receive – as a rule – fewer citations than publications in English.

**(5) Missing and/or incomplete databases in certain disciplines.** Bibliometric analyses can be poorly applied in certain disciplines (e.g., social sciences, humanities, computer science). The most important reason is that the literature from these disciplines are insufficiently covered in the major citation databases which focus on journal publications (Marx & Bornmann, 2015). For example, computer scientists tend to publish most of their work in conference proceedings rather than in journals. In other words, "bibliometric assessment of research performance is based on one central [but false] assumption: scientists, who have to say something important, do publish their findings vigorously in the open, international journal literature" (van Raan, 2008, p. 463).

Two additional problems with bibliometric indicators concern our understanding of numbers and of what information numbers can convey.

**(1) Poorly understood indicators**. Similarly to what happened with *p*-value and other inferential (e.g., Bayesian) statistics (e.g., Gigerenzer & Marewski, 2015), many of those relying on bibliometric indicators do not seem to know how to use them in meaningful ways, or what the numbers really mean. The JIF is a widely used indicator to infer "the impact" of single publications by a researcher. However, the indicator was originally developed to decide on the importance of holding journals in libraries. Paralleling how Fisher's and von Neyman and Pearson's respective statistical frameworks became alienated from their intended purposes (e.g., Gigerenzer, 2004; Gigerenzer et al., 2004; Gigerenzer & Marewski, 2015), the JIF was applied, with little conceptual fundament, to new tasks – judgments about the quality or relevance of scientific output. Similarly problematic, the *h*-index combines the output and citation impact of a researcher in a single number. However, with *h* papers having at least *h* citations each, the formula for combining both metrics is arbitrarily chosen: $h^2$ citations or $2h$ citations could have been used as well; just as $p > .06$ or $> .02$ could have defined statistical significance instead of the arbitrary .05 and .01.[3]

Until the end of the 20th century, bibliometrics was frequently conducted by experts in bibliometrics who knew about the typical problems with bibliometric studies, alongside with possible solutions. Since then "desktop scientometrics" (Katz & Hicks, 1997, p. 141) has become more and more popular. Here, research managers, administrators, and scientists from fields other than bibliometrics use "bibliometric data in a quick, unreliable manner" (Moed & Halevi, 2015, p. 1989). Bibliometric applications are sold by *Clarivate Analytics* (*InCites*) and *Elsevier* (*SciVal*), which provide ready-to-use productivity and impact indicators, foregoing available expertise and scrutiny from professional bibliometricians (Leydesdorff, Wouters, & Bornmann, 2016). As Retzer and Jurasinski (2009) point out – rightly – "a review of a scientist's performance based on citation analysis should always be accompanied by a critical evaluation of the analysis itself" (p. 395). Bibliometric applications, like *SciVal* or *InCites*, can deliver bibliometric results, but they cannot replace human judgement and expertise.

**(2) Impact is not equal to quality.** A citation-based indicator might capture (some) aspects of quality, but is not able to accurately *measure* quality –indicators are "largely built on sand", in the view of some (Macilwain, 2013, p. 255). According to Martin and Irvine (1983) citations reflect scientific impact as just one of three aspects of quality: correctness and importance are the other two. Applicability and real world relevance are further aspects scarcely reflected in citations. Moreover, revolutionary findings – those, indeed, leading, to scientific revolutions – are not necessarily highly cited ones. For example, Marx and Bornmann (2010) and Marx and Bornmann (2013) bibliometrically analyzed publications that were decisive in revolutionizing our thinking, such as those (1) that replaced the static view with Big Bang theory in cosmology, or (2) that dispensed with the prevailing fixist point of view (fixism) in favor of a dynamic view of the Earth where the continents move through the Earth's crust. As those bibliometric analyses show, several publications that propelled the transition from one theory to another are lowly cited.

---

[3] Indeed, we likely use .05 and .01 today, only because Fisher did not have any tables for other numbers – his archenemy, Karl Pearson, appeared to have refused to share those with him (Gigerenzer & Marewski, 2015).

What is more, in all areas of science, important publications might be recognized as such only many years after publication – these articles are like sleeping beauties, only that – unlike in fairy tales – nobody starts to kiss them (Ke, Ferrara, Radicchi, & Flammini, 2015; van Raan, 2004). The Shockley-Queisser paper (Shockley & Queisser, 1961) – describing the limited efficiency of solar cells on the basis of absorption and reemission processes – is one such sleeping beauty (Marx, 2014). This groundbreaking paper was rarely cited within the first 40 years after its publication. While initial lack of recognition can be compensated, in the long run, by exploding citation rates, the time, resources, and tenure-cases lost until those kisses, finally, produce themselves might be harder to make up for. Even worse, sometimes papers that are highly cited perpetuate factual mistakes, miconceptions, or misunderstandings contained in them. For instance, a highly cited paper by Preacher and Hayes (2008) recommends using a certain statistical procedure to test mediation. This procedure is widely used in the disciplines of psychology and management; however, this procedure actually produces biased statistical estimates because it ignores a key assumption made by the estimator (i.e., that the mediator is not endogenous; see Antonakis, Bendahan, Jacquart, & Lalive, 2010; Kline, 2015; Shaver, 2005; Smith, 2012).

**Science evaluation from a statistical point of view: Universal and automatic classifiers do not exist**

The problem with recognizing quality science is that – statistically-speaking – judgments about the quality of research (or people or institutions) represent classifications. As any classifier, also indicator-based classifications will never yield perfect results.[4] Instead, *correct positives* (giving "good research" good evaluations) and *correct negatives* (giving "bad research" bad evaluations), *false positives* (giving "bad research" good evaluations) and *false negatives* (giving "good research" bad evaluations) will occur (Bornmann & Daniel, 2010). To develop accurate classifiers, in areas other than science evaluation, one tests, out of sample, different classification (i.e., judgement) algorithms in model comparisons. This can be done, for example, in medical diagnosis (e.g., Marewski & Gigerenzer, 2012). Yet, in science evaluation that approach is not feasible for at least three reasons.

First, in order to determine classification (i.e., the judgements') accuracy one needs to have a criterion variable, which does not exist in research evaluation. That is, one does not really know for sure how good research is, given a valid outside yardstick. In certain areas of medical diagnosis, in contrast, it is relatively more straightforward to establish outside criteria. HIV, a cancer, or a bacterium might be present or not; if present and if the classifier predicts the medical condition to be present, one has a correct positive, if not present, and if the classifier predicts the condition to be present, one has a false positive, and so on.

Second, even if there were clear-cut, meaningful criteria in research evaluation, different operationalization of them can result in different classifiers coming out as winner in model comparisons. In any classification problem, what classifier maximizes classification accuracy depends on the criteria chosen, the predictor variables, the sample at hand, as well as other factors – there is simply no universally adequate classifier. Hence, there is also no universally "correct" way how to design a classfier, for instance, how to fix the size of the calibration and test samples needed to develop and test a classfier. But following the old ideals of universalism and automatism, many treat citation counts, JIFs, and other bibliometric variables as if they could be used as "classifiers" that were informative in all situations and their classifications independent of how different people used them.

---

[4] The various indicators (e.g., citation rates, JIF, *h*-index) alluded to above can be thought of predictor variables to be used in the "classification" of research output, people, or institutions. That said, even if it were possible to improve those predictor variables, the three problems listed below (lack of criterion variable, inexistence of universally accurate classifiers, unknown cost-benefit tradeoffs) still prevail. For instance, the suggestion to use citation percentiles instead of mean citation rates (Bornmann, 2013a) and to field-normalized citation counts (Waltman, 2016) does not solve those three problems. Much the same can be said with respect to other problems we did not even mention, including problems related to the interpretation of relevenat statistical relations, such as the notorious confusion of correlation and causation (Glänzel, 2010).

Third, when it comes to research evaluation the criterion one ought to be interested in is actually not just (likely unmeasurable) classification accuracy, but the costs and benefits associated with correct positives, correct negatives, false negatives, and false positives. What is the cost (e.g., to society, the taxpayer, the individual scientist) if just one brilliant, groundbreaking piece of work is classified as "bad", or if millions of trivial findings are classified as "good" (e.g., simply because they are cited)? In science evaluation and many other classification tasks, the "real" cost and benefit of classifications are hard to estimate or fully unknowable. Even in medicine, where costs and benefits are sometimes knowable, it is not always clear how they ought to be traded off against each other.[5]

Even if one worked with fully hypothetical costs-benefit structures, different stakeholders will place more or less importance on different costs and benefits (e.g., different scientists, evaluators, or politicians might have different agendas when it comes to evaluating the research of their colleagues, institutions, or themselves). Assumptions about cost-benefit structures will, too, differ across contexts (e.g., a false negative in cancer research is not the same as one in social psychology). But with each change in the assumed cost-benefit structure, the performance of each classifier can change. Statistically speaking, universal and automatical classifiers do not exist.

---

[5] To illustrate this point, the costs that come with a false-negative mammogram will *cleary* outweigh the costs come with a false-positive result, so all women should regularly participate in breast cancer screening, one might think. Indeed, doctors, lobbyists, politicians, and other professionals argue they should, and a billion dollar-generating business lives from national screening programs. Yet, out of 1,000 women, aged 50 years and older, who participate in breast cancer screening for 10 years on average, 100 will experience severe costs, including false alarms, biopsies or psychological distress, and 5 with non-progressive cancer will have unnecessary treatment (e.g., complete or partial breast removal). Of those 1,000 women, 4 will die from breast cancer, and 21 will die from any type of cancer. To compare, of those women who do not undergo breast cancer screening, 5 will die from breast cancer, and 21 from any type of cancer, and there will be no screening-related costs (e.g., unnecessary biopsies, treatment) (Gigerenzer, 2014). Now, is the reduction from 5 to 4 women who will die from breast cancer – which is, actually, the 25% risk reduction mentioned above – "worth" the costs of screening? Moreover, take opportunity costs into account: tax money poured into screening programs cannot be spent on other purposes, such as cancer prevention programs, traffic education, or environmental protection. Having numbers can help to see how difficult the problem is; but even so, numbers do not, on their own, solve what is a problem of judgment.

# What is the remedy? Understand that heuristics are tools for judgements under uncertainty, including for science evaluation

In the introduction to this essay, we wrote that numbers have been *turned* into the new opium of the people. We did not write that they *are* opium. Opium is a dangerous drug; numbers are not drugs, but they can cloud one's thinking and affect one's emotional state like drugs. Similar to many drugs, numbers can do not only harm but, when insightfully and diligently used, can actually be quite beneficial.

So let us be clear: we *do not* advocate to get rid of numbers in judgement. From experimentation to computer simulation, for most scientific judgments, numbers are indispensable – also when it comes to science evaluation. For instance, in playing the devil's advocate, one could ask why a numeric indicator for gauging the quality of publications is necessary; one could simply read the publications and assess their quality (Gigerenzer & Marewski, 2015). However, such a practice would not be possible in large-scaled evaluations of (competing) research units. Furthermore, large research projects are often transdisciplinary. Experts would be necessary to assess the quality of each single publication. If one did not want to trust bibliometric indicators, one would thus either have to recruit the experts, or forego large-scale evaluations in the first place.[6]

What we *do* advocate is that those who invoke numbers to assess the "quality" of research, people, or institutions do not pretend, at the same time, to also avoid the uncertainties that come with *any* judgment. Likewise, we advocate that those who dare to rely on their judgment are not automatically forced to make up numbers ("facts") even when doing so does not make any sense. In our view, the Enlightenment's (i.e., Kant's) *Sapere audere!* – dare to know – ought to come accompanied with a *Invite good human judgment!* How can this become common practice in science evaluation?

---

[6] Another problem with expert judgments concerns communication: As Hug, Ochsner, and Daniel (2014) point out, "scholars possess knowledge, which allows them to recognize quality research; however, they cannot articulate this knowledge clearly and easily" (p. 61). Polanyi (1967) calls this situation tacit knowing which refers to "the fact that we can know more than we can tell" (p. 4). Gigerenzer (2014) calls such inarticulate, unconscious reasons intuition and points out that many experts (e.g., in business, medicine, and beyond) feel compelled to invent numbers, facts, and arguments after the fact – only to justify their gut feelings to distrusting individuals or for accountability purposes.

Many believe, the more information available, the better. And if only little information is out there, one should, ideally, go out and seek more. Once all facts are on the table, one can then sit down and integrate them all in order find the best solution. This classic view, rooted in the Enlightenment's focus on analysis goes strong in many areas of psychology, economics, the information sciences, and beyond. Yet, science evaluation and the vast majority of other judgment tasks we face in our private and professional life do not resemble *gambles of chance*. In gambles of chance, all information is knowable or can be reliably estimated, and nothing unexpected will happen.

In dice-throwing, roulette, and card games, for example, it is possible to calculate the probabilities of different outcomes, and the consequences of those outcomes are known. Outcomes when throwing an unbiased dice are the numbers 1, 2, 3, 4, 5, and 6, with each outcome having an equal probability of occurring: 1/6 is the relative frequency in the long run. Similarly, when playing roulette, one knows what is at stake: as a consequence of an outcome, one can either lose or win the money on the table; or in the case of Russian roulette with two guns, the only four possible consequences are (i) a bullet in one's skull, (ii) a bullet in the opponent's skull, (iii) a bullet in both skulls, (iv) a bullet in no skull.

In science and everyday life, in contrast, surprises can occur and there is an unknown amount of unknown unknowns: Do we know the range of possible things that might happen to us tomorrow, next week, or in ten years from now? Do we know which technology will be invented tomorrow? Is it knowable how many scientists who are not "productive" enough by yearly publication numbers, will in five, ten, fifteen years from now, make *the* discovery that revolutionizes their field? Gigerenzer (e.g., 2014) uses the term *uncertainty* to refer to the common situation where not all outcomes, their consequences, and probabilities of occurrences are knowable. When they are knowable, one speaks of *risk*. Situations of risk invite for trying to find, through calculation or logic, best solutions to given judgment tasks. It is those rare situations of knowable risks, where more information and optimal information integration yields better judgments.[7]

Uncertainty, in turn, affords relying on simple rules of thumb that allow making good-enough judgments based on little information, so called *fast-and-frugal heuristics* (Gigerenzer et al., 1999). The fast-and-frugal heuristics research program underlines that people can make smart decisions, because they can adaptively draw from a repertoire of strategies, including multiple heuristics for judgments under uncertainty and other more complex strategies (e.g., Bayesian ones) for decisions under risk (see Gigerenzer & Gaissmaier, 2011; Katsikopoulos, 2011; Marewski, Gaissmaier, & Gigerenzer, 2010, for overviews). In contrast to those other strategies, heuristics purposefully ignore information, both simplying decision making and leading to (reasonably) fast judgments (hence frugal and fast). Importantly, under uncertainty, a trade-off between simplicity and accuracy is not necessarily incurred; to the contrary, ignoring information is often key to making good judgments, as numerous computer simulations and mathematical analyses suggest (e.g., Czerlinski, Gigerenzer , & Goldstein, 1999; Gigerenzer & Brighton, 2009). Selecting adaptively from a toolbox of judgment strategies implies that not all tools are used in all situations, neither are they useful universally. Instead, each of the different tools in the repertoire can help only in certain situations. Knowing *when* to rely on which tool is at the heart of the art of clever decision making.

For instance, why and when various fast-and-frugal heuristics can help making good judgments under uncertainty can be explored in terms of what is known, in statistics, as the *bias-variance dilemma* (Geman, Bienenstock, & Doursat, 1992; Gigerenzer & Brighton, 2009). Calculable risks, in turn, can afford for being tackled with seemingly automatic and universal quantitative procedures. Bayes theorem is just one of them; the maximization of (subjective expected) utility (e.g., Edwards, 1954), as it is also assumed to drive the egoistic homo economicus, is another. Calculable risks can also afford measuring how much actual human behavior deviates from normatively "optimal" (e.g., Bayesian) solutions – and, hence, for classifying behavior as suboptimal or irrational whenever that behavior does not live up to the calculated expectations (e.g., Gigerenzer & Gaissmaier, 2015).[8]

---

[7] Indeed, as the reader might have noticed, throughout this essay, we have never used the term risk; instead we have been referring to judgments under uncertainty from the start. The notions of risk and uncertainty in decision making as Gigerenzer (e.g., 2014; Mousavi & Gigerenzer, 2014) and we (e.g., Hafenbrädl, Waeger, Marewski, & Gigerenzer, 2016) use them have been developed based on the work of Knight (1921) and others (e.g., Savage, 1954) and can be distinguished from other (related) notions in different disciplines (see e.g., Bookstein, 1997, for an example from informetrics). Specifically, risk refers to situations in which probabilities can be reliably estimated or are known, such as the gambles of chance mentioned above (e.g., roulette, card games, lotteries) which represent predicable and well-defined problems. The notion of *small worlds*, put forward by the father of modern Bayesian decision theory, Savage (1954), too, refers to such settings of perfect information. Uncertainty, in turn, comes with unknown, unknowable, or mathematically imprecise probabilities – fuzzy situations that Binmore (2009) dubbed *large* (or uncertain) *worlds*.

Unfortunately, in the past century, too much social science research has, rather than investigating genuine tools for judgments under uncertainty and finding out when each of those tools works, vainly attempted to reduce uncertainty to risk (see Gigerenzer & Marewski, 2015; Gigerenzer & Murray, 1987).[9] As a result, many academics do not only seem to treat science evaluation and statistics as if there were just one type of universal tool available (e.g., *h*-index for all assessments of researchers or NHST for all statistical inference), but also the idea that simple, intuitive rules – folk wisdom of sorts, such as *Do not put your eggs into the same basket!, Split your resources equally!*, *Take the best!* (e.g., Gigerenzer, 2014; Gigerenzer et al., 1999) – could outwit detailed analysis in the first place seems counterintuitive to many.

In short, in many ways, the Enlightenment's ideals continue to go strong. But the ideals start breaking down when one accepts the world as what it really is: uncertain. Under uncertainty, heuristics can outperform more complex judgment procedures – be it when it comes to medical diagnosis, sports forecasting, financial investment, criminal investigations, or election polling to name but a few (e.g., Gigerenzer & Gaissmaier, 2011; Gigerenzer, Hertwig, & Pachur, 2011; Goldstein & Gigerenzer, 2009; Hafenbrädl et al., 2016). We think that heuristics could improve the uncertain world of science, too – namely (i) by transforming the workings of science as a social system and (ii) by aiding people's judgments about science.

---

[8] For instance, following the heuristics-and-biases framework mentioned above, behaviour that deviates from the maximization of subjective expected utility and other normative yardsticks (e.g., the rules of logic) is considered irrational, biased, or otherwise fallacious.

[9] Under uncertainty, as Gigerenzer and we (e.g., Hafenbrädl et al., 2016) use the term, surprises can occur: the decision maker does not know all options, their consequences, and the probabilities of those consequences occurring, and/or the available information does simply not allow to reliably estimate those. In such situations, the premises of rational (e.g., Bayesian) decision theory are not met, and classic (e.g., optimization) approaches to decision making and human rationality are rendered inappropriate. Traceable to the Enlightenment and to thinkers such as Blaise Pascal, Pierre Fermat, and Daniel and Nicholas Bernoulli, those classic approaches to decision making still dominate much thinking in science (e.g., notably economics, biology, psychology) and society, namely when decision making problems that entail uncertainty (or risk *and* uncertainty at the same time) are treated as if they just entailed (calculable) risk. In the social sciences, prominent examples and derivatives of those classic approaches are not only the above-mentioned maximization of (subjective) expected utility and Bayesian inference models (e.g., Arrow, 1966; Becker, 1993; Chater & Oaksford, 2008; Edwards, 1954; Savage, 1954; von Neumann & Morgenstern, 1947), but also widely used statistical tools (e.g.,linear models that estimate coefficients while minimizing errors).

Regarding the first point, imagine only how academia would look like if all researchers followed the simple heuristics to *only* use statistics in result sections that they fully comprehend? Regarding the first point, imagine only how academia would look like if all researchers followed the simple heuristics to *only* cite those papers they have really read and understood (see e.g., Penders, 2018), or to *only* use statistics in result sections that they fully comprehend? Such simple rule could be backed-up by others. An example is *Avoid silo-thinking; that is, also search for knowledge outside of your own discipline!* Often similar questions are asked in different disciplines, with similar notions being given different labels (see Arkes & Ayton, 1999; Marewski & Link, 2014, for examples concerning psychology and economics). A punchline for a simple rule for science evaluation would be to require science evaluators – much like along the lines of the *Harnack Principle* practiced at the Max Planck Society (Max Planck Society, 2018) – to actually read and reflect upon the papers of those they evaluate: *Only use citation-based indicators alongside expert judgements of papers, and those papers have to be carefully read and understood)!* A result of that rule would be that those evaluating others would have to be experts in the very same area – instead of generalists who can, at best, only judge the "quality" of work from the outside by relying on citation and publication numbers or other seemingly universal indicators. Hence, science administrators always would have to integrate expert judgements in evaluations.

As to the second point – to aid judgment about science – the difficulty and beauty of science evaluation lies in knowing when to use which indicator. Contrary to what many science evaluators do, who pretend to apply single indicators (e.g., *h*-indices) universally, there is no single indicator that can be meaningfully applied to all problems. There is also no indicator that will lend results automatically, that is, indepent of the person or institution using it. Moreover, in many situations, there will be no clear answer whether using a specific indicator is "right" vs. "wrong" or "optimal" vs. "suboptimal"; instead judgment is required and uncertainty will prevail.

Finally, one should not always prefer advanced indicators over simple ones and vice versa. For instance, citation counts and other simple indicators can be helpful when considered within a single scientific field, such that there is no need for field normalization (see footnote 4); particularly citation counts are easy to understand and therefore it is relatively easy for end-users to recognize the problems and limitations that come with citation count-based judgments (Waltman & van Eck, 2016).

We like to think of science evaluation in terms of judgments under uncertainty that can be best tackled if one is prepared to choose – with all uncertainty – from a repertoire of different techniques – including different indicator and non-indicator based ones. The challenge consists of selecting techniques that yield satisfactory results. Nobel laurate Herbert Simon (e.g., 1956, 1990) developed the notion of *satisficing*: the idea is not to aspire identifying the optimal option from a set, but one that is sufficiently good to meet the goal. Aspiring to select sufficiently good heuristics, and knowing that, in so doing "mistakes" are inevitable, is what science evaluation under uncertainty ought to be about. That view corresponds to the fast-and-frugal heuristics approach to decision making introduced above. It also resembles the *toolbox view of statistitics* (e.g., Gigerenzer & Marewski, 2015): rather than pretending that the statistical toolbox contained just one type procedure for making statistical inferences – be it NHST for frequentists or Bayesian procedures for Bayesians – meaingful statistics can be best conceived of in terms of a repertoire of different statistical tools which many of them being useful in quite different situations.[10]

**Let expertise in the research field and expertise in bibliometrics inform the selection of heuristics from the toolbox**

Consistent with a toolbox view on judgment and statistics, expertise can aid decision making, namely when it comes to deciding which of the various tools to rely upon. At least two kinds of expertise are necessary if metrics are used by administrators.

First, judgments made by a field's experts are necessary to interpret the numbers in a bibliometric report and to place them in the institutional and field-specific context. When conducting a bibliometric evaluation of a scientist, journal, or an institution, judgments should be made within the field, and not by people outside of that area of research. This is what the notion of *informed peer review* is about: "the idea [is] that the judicious application of specific bibliometric data and indicators may inform the process of peer review, depending on the exact goal and context of the assessment. Informed peer review is in principle relevant for all types of peer review and at all levels of aggregation" (Wilsdon et al., 2015, p. 64). Heuristics are – much like science evaluation *should be* – context depedent, and knowing the context well can aid to select satisfactory heuristics.

Second, expertise in a field ought to be combined with expertise in bibliometrics. Much like a non-physician or a non-pilot should not attempt to diagnose patients or fly airplanes even if convenient diagnosis tools are sold on the internet or if flight simulation software is readily available, staff with insufficient bibliometric expertise (e.g., administrators) should not be put in positions where those non-experts have to assess units (e.g. scientists or institutions) – even if freely available bibliometric tools such as *InCites* or *SciVal* seemingly make those tasks simple.[11]

---

[10] NHST, for instance, is useful in situations when there is little knowledge about the phenomenon to tbe studied and precise competing hypotheses cannot be formulated. Using Bayes' rule makes sense when priors can be reliably estimated. Yet other situations may warrant completely different tools – ranging from exploratory data analysis to out-of-sample tests of competing custom-built models, exhausitive graphical data analysis, or the simplest of descriptive statistics.

Professional bibliometricians do not only have access to a repertoire of different databases and indicators and are used to choosing among their tools, but they may also advise the client against a bibliometric report in cases where bibliometrics can be scarcely applied (e.g. in the humanities) or point to other problems alongside with possible solutions. Hicks et al. (2015) formulated 10 principles – the *Leiden Manifesto* – which guide experts in the field of scientometrics (see also Bornmann & Haunschild, 2016). For example, performance should be measured against the research missions of the institution, group or researcher (principle no. 2) or the variation by field in publication and citation practices should be considered by using field-normalized citation scores in cross-field evaluations (principle no. 6). Thus, if a report is commanded from bibliometricians, the responsible administrator should try to understand the report by discussing it with the bibliometricians.

**Become statistically literate in order to evaluate and communicate about different tools for science evaluation**

A toolbox approach to science evaluation calls for more than just expertise in bibliometrics and in the domain of research. Any individual involved with research evaluation ought to increase her knowledge on statistics and data analysis techniques. It is, at the end of the day, the comprehension of numbers that will help administrators to discuss bibliometric reports with the bibliometricians and scientists from the field, and it is the comprehension of numbers that will aid administrators to see the bibliometric report's limitations. Similarly, it is statistical knowledge that allows playing the devil's advocate on numbers provided by others or compiled by oneself. Notably, social scientists should know how to program computer simulations to scrutinize their own judgments and classifications. Just imagine what would happen if everybody followed the simple rule of thumb to *only* rely on numbers (e.g., Bayes factors, *p*-values, bibliometric indicators) and quantitative models (e.g., regressions, classification trees) they are able to fully program from scratch themselves, that is, without any help of off-the-shelf software (e.g., *SPSS*) and without making use of their pre-programmed functions? Likely there would be all five: more reading, more thinking, more informed discussions about contents, less uninformed abuse of numbers, and more experts in bibliometrics and statistics.

---

[11] Thor, Marx, Leydesdorff, and Bornmann (2016) introduced the program CRExplorer (see www.crexplorer.net) which forces the application of bibliometrics by experts in the field of evaluation. The user of the program has to select and import the publications on which citation impact is measured. The program counts the cited references in these publications to identify the most influential works. Without a carefully selected publication set as import—which can only be done with the necessary expertise—the results of the program become meaningless.

Scientists and science evaluators ought to be familiar with basic statistical principles for statistical reasoning under uncertainty, such as testing one's predictions on data that differs from the data that served the development of the predictions, that is, out of sample or out of population (e.g., Marewski & Olsson, 2009; Pitt, Myung, & Zhang, 2002; Roberts & Pashler, 2000). To give another example, evaluators should be familiar with different techniques of exploring, representing, visualizing, and communicating data and results, ranging from effect size displays to *natural frequencies* (Hoffrage, Lindsey, Hertwig, & Gigerenzer, 2000).[12] Seeing the same numbers presented in different ways is key to understanding them. Indeed, in basic research on fast-and-frugal heuristics and statistics, repertoires of different judgmental and statistical "tools" are not simply "invented", but instead the performance of different "tools" is investigated in extensive mathematical analyses and/or computer simulations. By analogy, science evaluators ought to be sufficiently familiar with data analysis techniques in order to be able to evaluate their own tools for science evaluation. For instance, evaluators ought to understand the parallels between science evaluations and any statistical classification problem, as described in the section „Science evaluation from a statistical point of view". Science Evaluators ought to know the statistics helpful for evaluating classifiers (e.g., sensitivity, specificity, receiver-operating curves), and how to develop and test different classifiers, including heuristic ones (e.g., fast-and-frugal trees; Luan, Schooler, & Gigerenzer, 2011; Martignon, Katsikopoulos, & Woike, 2008) in cross validation.

---

[12] Natural frequencies, for instance, offer a transparent format to describe the outcomes of classifications; less transparent formats are fractions and percentages. Take breast cancer screening as example (see Gigerenzer, Gaissmaier, Kurz-Milcke, Schwartz, & Woloshin, 2007). What is the likelihood that a woman, who gets a positive mammogram, has cancer? The probability is about 10%, if the breast cancer screening mammogram has a sensitivity of 90%, and a false-positive rate of 9%, with the prevalence of breast cancer being 1%. Confused? Here, are the same statistics, this time represented as natural frequencies. Imagine 1,000 women. A prevalence of 1% means that 10 of 1,000 women will have breast cancer and 990 not. A sensitivity of 90% means that of those 10 women with cancer, 9 will have a positive mammogram. A false positive rate of 9% means that of the 990 women without cancer about 89 will nevertheless have a positive mammogram. Hence, a total of 98 women (= 9+89) will test positive, but of these 98 only 9 will have cancer, which corresponds to roughly 10%.

In short, science evaluation – like most judgments in our daily lives – concerns uncertainty; however, in many educational curricula around the world, more emphasis seems to be placed on mathematics of certainty (e.g., calculus) than on statistical thinking and genuine tools for dealing with uncertainty, such as heuristics (see Gigerenzer, 2014). The best bet would be that statistical literacy and the critical reflection upon statistics form part of school education, much like reading, writing, or history. As a helpful side effect, then also those putting stakes on bibliometric and other numbers – including mere consumers of science evaluations such as administrators, politicians or normal citizens – might have better intuitions about what to take from those statistics and what not, as well as when to not request numbers in the first place.

**A note on how to put the toolbox into practice**

We do not advocate for any of our recommendations to be put in practice in isolation. Rather, we believe that it is the simultenous implantation of all three that can aid science evaluation. To illustrate this, imagine only the recommendation that expert bibliometricians perform evaluations would be implemented. The result could be that those "experts" in bibliometric numbers become too powerful in a research evaluation: decision making may then be based too much on technical bibliometric criteria rather than on substantive (including qualitative) considerations made by researchers from the field. In contrast, if those who perform the evaluation are not only experts in bibliometrics but also experts in the field under study, and if moreover those experts understand that they are, essentially, making decisions under uncertainty and that, here, often simple decision strategies can help (e.g., Mousavi & Gigerenzer, 2014), then there might be room for both quantitative evaluation and substantive, qualitative considerations. Finally, if all involved have good statistical knowledge then they might additionally be in a better position to know, for instance, when which of many different quantitative indicators is useful, or how to best collect and evaluate general publication statistics for a given field.

Importantly, an implication of the toolbox view is the need of establishing an error culture – designs of institutions, work contracts, and review proceedures that aid dealing with the mistakes that any judgement under uncertainty can entail, including evaluations of people, departements, or research output. That does not necessarily only entail bolstering the effects of errors, but it also entails *investigating* the potential sources of errors in order to avoid them in the future. For instance, the results of Franzoni, Scellato, and Stephan (2011) show that "cash incentives appear to encourage submission of research regardless of quality, as suggested by correlation with lower acceptance rates" (p. 703). Ideally, one would get rid of those incentives, or if that is not possible, create and systematically *test* counterbalancing ones, such as requiring experts in the research field to establish whether a publication is "worthwhile" to be counted.

Note the words "investigating" and "test" were put in italics in the previous sentences: steady empirical research is required to continuously evaluate the effects of science evaluations, including how they change people's behavior and science as a social system (Rijcke, Wouters, Rushforth, Franssen, & Hammarfelt, 2016). In a sense, it is ironic that those who advocate systematic metrics-based science evaluation and monitoring do not advocate, with the same fervor, a ceaseless systematic empirical evaluation of science evaluation itself. Here, perhaps the simple rule to avoid silo-thinking might help: an area – outside science – where sources of errors and counter-measures are systematically investigated and monitored is the international aviation industry; another area to turn to might be decision scientists' and behavioral economists' research on how incentives shape behavior.

## Conclusions

Substance abuse is not only potentiated by anxiety, but abuse also potentiates anxiety. People start to consume drugs because they are afraid; the more often they consume, the more frequently they will feel anxious – a vicious circle. Anxiety is, too, a driving factor behind phobias and obsessive compulsive disorders: Compulsory behaviors (e.g., ceaseless counting), intrusive thoughts (e.g., repeating magic numbers), or avoidance of stimuli and thoughts (e.g., about death) are born out of the desire to cope with unbearable fears, for example, in desperate attempts to gain control of an uncertain and potentially threatening future. Similar to drug abuse, the more often those behaviors are carried out, the more impetus they gain – upon executing a thought ritual or a compulsion, the fear momentarily decays away, but only to emerge ever more strongly thereafter. So the behavior is carried out again – a spiral of increasing addiction starts. One that – akin to a virus – can "infect" other individuals too. In the process, trust in rituals eventually displaces trust in one's own good judgments (Morschitzky, 2009).

It looks as if parts of academia and society suffer from collective anxiety, with various forms of number-crunching having developed into fear-reducing "mental drugs" to cope with the unreducible and unbearable uncertainties surrounding scientific judgment problems. Be it when it comes to evaluating theories, findings, or data, or when it comes to judging people, papers, and institutions, often there simply is no clear answer. But there are ignorant reviewers, hiring committee members, peers, lawyers, politicians and other respect-inspiring authorities who insist on obtaining a clear, "objective", justifiable answer. If the answer is not given by yourself, dear scientist, then another colleague, likely drugged by his or other's numbers, will be quick to offer one. This offer will then be marketed – at least initially – as implying progress, for instance, over your self-admitted ignorance or the pasts's mere intuitions and hunches.

Indeed, it is, ultimately, thanks to numbers, precise measurement, elaborated statistics, and quantification, that we have gone beyond Epicure's reasoning about atoms, the geographic drawings of the Roman Tabula Peutingeriana, shaman cures of diseases, or Newton's classical mechanics, all of which, so one might think, are located somewhere below us on a gigantic staircase parting from the past, reaching into the present and rising up into an even brighter future. We agree with the view that numbers aid science and society to progress. Yet, we also think that one should, perhaps, once in a while, walk the staircase of scientific and societal developments backward, in quest of what seems to have gotten lost on the way up: good human judgment. How can this be done? Thinking about the past can aid us to intuit, and hence possibly shape, the future. This is not only true for us, as human beings – we all are the past's future skulls – but also for the historical trends which might put the contemporary number-obsession into prespective.

## Author Note

This essay is meant to be (thought) provocative. Its goal is not to damnify numbers, but rather to trigger reflection on what numbers tend to do to science and society– a word of caution in the age of digitalization in science evaluation, written by two number-crunchers. One of us is a professional bibliometrician in one of Europe's largest research institution, with largest being measured in terms of institutes, employees, and annual number of publications. The other is a decision scientist who has dedicated his career to the building of mathematical and computational models of judgment. Numbers are our daily business.

## Acknowledgements

We would like to thank Justin Olds and Guido Palazzo for helpful comments on an earlier version of this manuscript.

# References

Aitkenhead, D. (2013). Peter Higgs: I wouldn't be productive enough for today's academic system. Retrieved July 5, 2016, from https://www.theguardian.com/science/2013/dec/06/peter-higgs-boson-academic-system

Ajiferuke, I., & Famoye, F. (2015). Modelling count response variables in informetric studies: Comparison among count, linear, and lognormal regression models. *Journal of Informetrics, 9*(3), 499-513. doi: 10.1016/j.joi.2015.05.001.

American Psychological Association. (1974). *Publication manual of the American Psychological Association* (2. ed.). Baltimore, MD: Garamond/Pridemark.

Anderson, J. R. (2007). *How can the human mind occur in the physical universe?* New York: Oxford University Press.

Antonakis, J., Bendahan, S., Jacquart, P., & Lalive, R. (2010). On making causal claims: A review and recommendations. *Leadership Quarterly, 21*(6), 1086-1120. doi: 10.1016/j.leaqua.2010.10.010.

Arbuthnott, J. (1710). An Argument for Divine Providence, Taken from the Constant Regularity Observ'd in the Births of Both Sexes. By Dr. John Arbuthnott, Physitian in Ordinary to Her Majesty, and Fellow of the College of Physitians and the Royal Society. *Philosophical Transactions, 27*(325-336), 186-190. doi: 10.1098/rstl.1710.0011.

Arkes, H. R., & Ayton, P. (1999). The sunk cost and Concorde effects: Are humans less rational than lower animals? *Psychological Bulletin, 125*(5), 591-600. doi: 10.1037/0033-2909.125.5.591.

Arrow, K. J. (1966). Exposition of the Theory of Choice under Uncertainty. *Synthese, 16*(3-4), 253-269. doi: 10.1007/Bf00485082.

Becker, G., S. (1976). *The economic approach to human behavior*. Chicago and London: The University of Chicago Press.

Becker, G. S. (1993). Nobel Lecture - the Economic Way of Looking at Behavior. *Journal of Political Economy, 101*(3), 385-409. doi: 10.1086/261880.

Biesenbender, S., & Hornbostel, S. (2016). The Research Core Dataset for the German science system: challenges, processes and principles of a contested standardization project. *Scientometrics, 106*(2), 837-847. doi: 10.1007/s11192-015-1816-y.

Binmore, K. (2009). *Rational decisions*. Princeton, NJ: Princeton University Press.

Bookstein, A. (1997). Informetric distributions .3. Ambiguity and randomness. *Journal of the American Society for Information Science, 48*(1), 2-10.

Bornmann, L. (2013a). How to analyse percentile citation impact data meaningfully in bibliometrics: The statistical analysis of distributions, percentile rank classes and top-cited papers. *Journal of the American Society for Information Science and Technology, 64*(3), 587-595.

Bornmann, L. (2013b). Research misconduct—definitions, manifestations and extent. *Publications, 1*(3), 87-98.

Bornmann, L. (2017). Measuring impact in research evaluations: A thorough discussion of methods for, effects of, and problems with impact measurements. *Higher Education, 73*(5), 775-787.

Bornmann, L., & Bauer, J. (2015). Which of the world's institutions employ the most highly cited researchers? An analysis of the data from highlycited.com. *Journal of the Association for Information Science and Technology, 66*(10), 2146-2148. doi: 10.1002/asi.23396.

Bornmann, L., & Daniel, H.-D. (2010). The Usefulness of Peer Review for Selecting Manuscripts for Publication: A Utility Analysis Taking as an Example a High-Impact Journal. *PLoS ONE, 5*(6), e11344. doi: 10.1371/journal.pone.0011344.

Bornmann, L., & Haunschild, R. (2016). To what extent does the Leiden Manifesto also apply to altmetrics? A discussion of the manifesto against the background of research into altmetrics. *Online Information Review, 40*(4), 529-543.

Bourguet, M.-N. (1987). Decrire, compter, calculer: The debate over statistics during the Napoleonic period. In L. Krüger, L. Daston & M. Heidelberger (Eds.), *The probalistic revolution* (Vol. 1, pp. 305-316). Cambridge, MA, USA: MIT Press.

Bowman, A. (2013). *Life and Letters from the Roman Frontier*: Taylor & Francis.

Chater, N., & Oaksford, M. (2008). *The probabilistic mind: Prospects for Bayesian cognitive science*. New York, NY: Oxford University Press.

Czerlinski, J., Gigerenzer, G., & Goldstein, D. G. (1999). How good are simple heuristics? In G. Gigerenzer, P. M. Todd & ABC Research Group (Eds.), *Simple heuristics that make us smart* (pp. 97- 118). New York: Oxford University Press.

de Bellis, N. (2009). *Bibliometrics and citation analysis: from the Science Citation Index to Cybermetrics*. Lanham, MD, USA: Scarecrow Press.

Edwards, W. (1954). The Theory of Decision Making. *Psychological Bulletin, 51*(4), 380-417. doi: 10.1037/h0053870.

European Research Council. (2012). *Guide for ERC Grant Holders* Brussels, Belgium: European Research Council (ERC).

Farber, D. (2013). Odds of dying from an asteroid strike: 1 in 74,817,414. Retrieved February 8, 2018, from https://www.cnet.com/news/odds-of-dying-from-an-asteroid-strike-1-in-74817414/

Forster, M. R. (2000). Key concepts in model selection: Performance and generalizability. *Journal of Mathematical Psychology, 44*(1), 205-231. doi: DOI 10.1006/jmps.1999.1284.

Franzoni, C., Scellato, G., & Stephan, P. (2011). Changing incentives to publish. *Science, 333*(6043), 702-703. doi: 10.1126/science.1197286.

Gelman, A., & Hennig, C. (2017). Beyond subjective and objective in statistics. *Journal of the Royal Statistical Society: Series A (Statistics in Society), 180*(4), 967-1033. doi: 10.1111/rssa.12276.

Geman, S., Bienenstock, E., & Doursat, R. (1992). Neural Networks and the Bias Variance Dilemma. *Neural Computation, 4*(1), 1-58. doi: DOI 10.1162/neco.1992.4.1.1.

Gigerenzer, G. (1987). Probabilistic thinking and the fight against subjectivity. In L. Krüger, G. Gigerenzer & M. Morgan (Eds.), *The probabilistic revolution: Ideas in the sciences* (Vol. 2, pp. 11-33). Cambridge, MA, USA: MIT Press.

Gigerenzer, G. (2004). Mindless statistics. *The Journal of Socio-Economics, 33*, 587-606.

Gigerenzer, G. (2008). *Rationality for Mortals: How People Cope with Uncertainty*. Oxford, UK: Oxford University Press.

Gigerenzer, G. (2014). *Risk savvy: How to make good decisions*. New York, NY, USA: Viking.

Gigerenzer, G., & Brighton, H. (2009). Homo Heuristicus: Why Biased Minds Make Better Inferences. *Topics in Cognitive Science, 1*(1), 107-143. doi: 10.1111/j.1756-8765.2008.01006.x.

Gigerenzer, G., & Gaissmaier, W. (2011). Heuristic decision making. *Annual Review of Psychology, 62*, 451-482.

Gigerenzer, G., & Gaissmaier, W. (2015). Decision making: Nonrational theories. In J. D. Wright (Ed.), *International encyclopedia of the social & behavioral sciences* (2. ed., pp. 911-916). Oxford: Elsevier.

Gigerenzer, G., Gaissmaier, W., Kurz-Milcke, E., Schwartz, L. M., & Woloshin, S. (2007). Helping Doctors and Patients Make Sense of Health Statistics. *Psychological Science in the Public Interest, 8*(2), 53-96. doi: 10.1111/j.1539-6053.2008.00033.x.

Gigerenzer, G., Gaissmaier, W., Kurz-Milcke, E., Schwartz, L. M., & Woloshin, S. (2008). Helping Doctors and Patients Make Sense of Health Statistics. *Psychological Science in the Public Interest, 8*(2), 53-96.

Gigerenzer, G., Hell, W., & Blank, H. (1988). Presentation and Content - the Use of Base Rates as a Continuous Variable. *Journal of Experimental Psychology-Human Perception and Performance, 14*(3), 513-525. doi: 10.1037/0096-1523.14.3.513.

Gigerenzer, G., Hertwig, R., & Pachur, T. (Eds.). (2011). *Heuristics: The foundations of adaptive behavior*. Oxford, UK: Oxford University Press.

Gigerenzer, G., Krauss, S., & Vitouch, O. (2004). The null ritual: what you always wanted to know about significance testing but were afraid to ask. In D. Kaplan (Ed.), *Handbook on quantitative methods in the social sciences* (pp. 391-408). Thousand Oaks, CA, UK: Sage.

Gigerenzer, G., & Marewski, J. N. (2015). Surrogate Science: The Idol of a Universal Method for Scientific Inference. *Journal of Management, 41*(2), 421–440. doi: 10.1177/0149206314547522.

Gigerenzer, G., & Murray, D. J. (1987). *Cognition as intuitive statistics*. Hillsdale, NJ: Lawrence Erlbaum.

Gigerenzer, G., Swijtink, Z., Porter, T., Daston, L. J., Beatty, J., & Krueger, L. (1989). *The empire of chance: How probability changed science and everyday life*. Cambridge, UK: Cambridge University Press.

Gigerenzer, G., Todd, P. M., & ABC Research Group (Eds.). (1999). *Simple heuristics that make us smart*. Oxford, UK: Oxford University Press.

Glänzel, W. (2010). On reliability and robustness of scientometrics indicators based on stochastic models. An evidence-based opinion paper. *Journal of Informetrics, 4*(3), 313-319. doi: 10.1016/j.joi.2010.01.005.

Glänzel, W., & Moed, H. (2013). Opinion paper: thoughts and facts on bibliometric indicators. *Scientometrics, 96*(1), 381-394. doi: 10.1007/s11192-012-0898-z.

Goldstein, D. G., & Gigerenzer, G. (2009). Fast and frugal forecasting. *International Journal of Forecasting, 25*(4), 760-772. doi: 10.1016/j.ijforecast.2009.05.010.

Guilford, J. P. (1942). *Fundamental Statistics in Psychology and Education* (3rd ed., 1956; 6th ed., 1978 with Fruchter, B. ed.). New York, NY, USA: McGraw-Hill.

Hacking, I. (1990). *The Taming of Chance*. Cambridge, UK: Cambridge University Press.

Hafenbrädl, S., Waeger, D., Marewski, J. N., & Gigerenzer, G. (2016). Applied Decision Making With Fast-and-Frugal Heuristics. *Journal of Applied Research in Memory and Cognition, 5*(2), 215-231. doi: 10.1016/j.jarmac.2016.04.011.

Hazelkorn, E. (2011). *Rankings and the reshaping of higher education. The battle for world-class excellence*. New York, NY, USA: Palgrave Macmillan.

Helbing, D. (2015a). *The Automation of Society is Next: How to Survive the Digital Revolution*: CreateSpace Independent Publishing.

Helbing, D. (2015b). *Thinking Ahead - Essays on Big Data, Digital Revolution, and Participatory Market Society*. Heidelberg, Germany: Springer.

Helbing, D., Frey, B. S., Gigerenzer, G., Hafen, E., Hagner, M., Hofsteter, Y., . . . Zwitter, A. (2017). Will democracy survive big data and artificial intelligence? Retrieved February 8, 2018, from https://www.scientificamerican.com/article/will-democracy-survive-big-data-and-artificial-intelligence/

Helbing, D., & Pournaras, E. (2015). Build digital democracy. *Nature, 527*(7576), 33-34.

Hicks, D., Wouters, P., Waltman, L., de Rijcke, S., & Rafols, I. (2015). Bibliometrics: The Leiden Manifesto for research metrics. *Nature, 520*(7548), 429-431.

Hoffrage, U., Lindsey, S., Hertwig, R., & Gigerenzer, G. (2000). Medicine - Communicating statistical information. *Science, 290*(5500), 2261-2262. doi: 10.1126/science.290.5500.2261.

Hoffrage, U., & Marewski, J. N. (2015). Unveiling the Lady in Black: Modeling and aiding intuition. *Journal of Applied Research in Memory and Cognition, 4*(3), 145-163. doi: 10.1016/j.jarmac.2015.08.001.

Holton, G., Chang, H., & Jurkowitz, E. (1996). How a scientific discovery is made: a case history. *American Scientist, 84*(4), 364-375.

Hug, S. E., Ochsner, M., & Daniel, H.-D. (2014). A framework to explore and develop criteria for assessing research quality in the humanities. *International Journal for Education Law and Policy, 10*(1), 55-64.

Hvistendahl, M. (2013). China's publication bazaar. *Science, 342*(6162), 1035-1039. doi: 10.1126/science.342.6162.1035.

Kahnemann, D., Slovic, P., & Tversky, A. (1982). *Judgment under uncertainty: heuristics and biases*. New York, USA: Cambridge University Press.

Katsikopoulos, K. V. (2011). Psychological Heuristics for Making Inferences: Definition, Performance, and the Emerging Theory and Practice. *Decision Analysis, 8*(1), 10-29. doi: 10.1287/deca.1100.0191.

Katz, J. S., & Hicks, D. (1997). Desktop scientometrics. *Scientometrics, 38*(1), 141-153. doi: 10.1007/bf02461128.

Ke, Q., Ferrara, E., Radicchi, F., & Flammini, A. (2015). Defining and identifying Sleeping Beauties in science. *Proceedings of the National Academy of Sciences, 112*(24), 7426–7431. doi: 10.1073/pnas.1424329112.

Kline, R. B. (2015). The Mediation Myth. *Basic and Applied Social Psychology, 37*(4), 202-213. doi: 10.1080/01973533.2015.1049349.

Knight, F. H. (1921). *Risk, Uncertainty and Profit*. New York, NY: Houghton Mifflin.

Kostoff, R. N. (1998). The use and misuse of citation analysis in research evaluation. *Scientometrics, 43*(1), 27-43.

Kruschke, J. K. (2010). Bayesian data analysis. *WIREs Cogn Sci, 1*, 658-676. doi: 10.1002/wcs.72.

Kuhn, T. S. (1962). *The structure of scientific revolutions* (2. ed.). Chicago, IL, USA: University of Chicago Press.

Leibniz, G. W. (1951). Toward a universal characteristic. In P. P. Wiener (Ed.), *Selections* (pp. 17-25). New York, NY, USA: Scribner's (Original work published 1677).

Leydesdorff, L., Wouters, P., & Bornmann, L. (2016). Professional and citizen bibliometrics: complementarities and ambivalences in the development and use of indicators—a state-of-the-art report. *Scientometrics, 109*(3), 2129–2150. doi: 10.1007/s11192-016-2150-8.

Lindley, D. V. (1983). Theory and Practice of Bayesian Statistics. *Statistician, 32*(1-2), 1-11. doi: Doi 10.2307/2987587.

Lindley, D. V. (1986). Comment. *American Statistician, 40*(1), 6-7. doi: Doi 10.2307/2683107.

Luan, S. H., Schooler, L. J., & Gigerenzer, G. (2011). A Signal-Detection Analysis of Fast-and-Frugal Trees. *Psychological Review, 118*(2), 316-338. doi: 10.1037/a0022684.

Macilwain, C. (2013). Halt the avalanche of performance metrics. *Nature, 500*(7462), 255.

Marewski, J. N., Gaissmaier, W., & Gigerenzer, G. (2010). Good judgments do not require complex cognition. *Cognitive Processing, 11*(2), 103-121. doi: 10.1007/s10339-009-0337-0.

Marewski, J. N., & Gigerenzer, G. (2012). Heuristic decision making in medicine. *Dialogues in Clinical Neuroscience, 14*(1), 77-89.

Marewski, J. N., & Link, D. (2014). Strategy selection: An introduction to the modeling challenge. *Wiley Interdisciplinary Reviews-Cognitive Science, 5*(1), 39-59. doi: 10.1002/wcs.1265.

Marewski, J. N., & Olsson, H. (2009). Beyond the Null Ritual Formal Modeling of Psychological Processes. *Zeitschrift Fur Psychologie-Journal of Psychology, 217*(1), 49-60. doi: 10.1027/0044-3409.217.1.49.

Martignon, L., Katsikopoulos, K. V., & Woike, J. K. (2008). Categorization with limited resources: A family of simple heuristics. *Journal of Mathematical Psychology, 52*(6), 352-361. doi: 10.1016/j.jmp.2008.04.003.

Martin, B. R., & Irvine, J. (1983). Assessing basic research - some partial indicators of scientific progress in radio astronomy. *Research Policy, 12*(2), 61-90.

Marx, W. (2014). The Shockley-Queisser paper – A notable example of a scientific sleeping beauty. *Annalen der Physik, 526*(5-6), A41-A45. doi: 10.1002/andp.201400806.

Marx, W., & Bornmann, L. (2010). How accurately does Thomas Kuhn's model of paradigm change describe the transition from a static to a dynamic universe in cosmology? A historical reconstruction and citation analysis. *Scientometrics 84*(2), 441-464.

Marx, W., & Bornmann, L. (2013). The emergence of plate tectonics and the Kuhnian model of paradigm shift: a bibliometric case study based on the Anna Karenina principle. *Scientometrics, 94*(2), 595-614. doi: 10.1007/s11192-012-0741-6.

Marx, W., & Bornmann, L. (2015). On the causes of subject-specific citation rates in Web of Science. *Scientometrics, 102*(2), 1823-1827.

Max Planck Society. (2018). The Max Planck Approach. Retrieved February 8, 2018, from https://www.mpg.de/39586/MPG_Introduction

McCloskey, D. N., & Ziliak, S. T. (1996). The Standard Error of Regression. *Journal of Economic Literature, XXXIV*(March), 97-114.

Moed, H. F., & Halevi, G. (2015). The Multidimensional Assessment of Scholarly Research Impact. *Journal of the American Society for Information Science and Technology, 66*(10), 2330-1643.

Morschitzky, H. (2009). *Angststörungen: Diagnostik, Konzepte, Therapie, Selbsthilfe*. Wien: Springer.

Mousavi, S., & Gigerenzer, G. (2014). Risk, uncertainty, and heuristics. *Journal of Business Research, 67*(8), 1671-1678.

Paulus, F. M., Rademacher, L., Schäfer, T. A. J., Müller-Pinzler, L., & Krach, S. (2015). Journal Impact Factor Shapes Scientists? Reward Signal in the Prospect of Publication. *PLoS ONE, 10*(11), e0142537. doi: 10.1371/journal.pone.0142537.

Penders, B. (2018). Ten simple rules for responsible referencing. *PLOS Computational Biology, 14*(4), e1006036. doi: 10.1371/journal.pcbi.1006036.

Pitt, M. A., Myung, I. J., & Zhang, S. B. (2002). Toward a method of selecting among computational models of cognition. *Psychological Review, 109*(3), 472-491. doi: 10.1037//0033-295x.109.3.472.

Polanyi, M. (1967). *The tacit dimension*. London, UK: Routledge & Kegan Paul.

Preacher, K. J., & Hayes, A. F. (2008). Asymptotic and resampling strategies for assessing and comparing indirect effects in multiple mediator models. *Behavior Research Methods, 40*(3), 879-891. doi: Doi 10.3758/Brm.40.3.879.

Reich, E. S. (2013). Science publishing: The golden club. *Nature, 502*(7471), 291-293.

Research Excellence Framework. (2014a). About the REF. Retrieved February 8, 2018, from http://www.ref.ac.uk/2014/about/

Research Excellence Framework. (2014b). Research Excellence Framework. Retrieved February 8, 2018, from http://www.ref.ac.uk/2014/

Research Excellence Framework. (2018). About the REF. Retrieved February 8, 2018, from http://www.ref.ac.uk/about

Retzer, V., & Jurasinski, G. (2009). Towards objectivity in research evaluation using bibliometric indicators: a protocol for incorporating complexity. *Basic and Applied Ecology, 10*(5), 393-400. doi: 10.1016/j.baae.2008.09.001.

Rijcke, S. d., Wouters, P. F., Rushforth, A. D., Franssen, T. P., & Hammarfelt, B. (2016). Evaluation practices and effects of indicator use—a literature review. *Research Evaluation, 25*(2), 161-169. doi: 10.1093/reseval/rvv038.

Roberts, S., & Pashler, H. (2000). How persuasive is a good fit? A comment on theory testing. *Psychological Review, 107*(2), 358-367.

Savage, L. J. (1954). *The foundation of statistics*. New York, NY: Wiley.

Schatz, G. (2014). The faces of Big Science. *Nature Reviews Molecular Cell Biology, 15*(6), 423-426. doi: 10.1038/nrm3807.

Scheidel, W. (1994). Libitina's bitter gains: seasonal mortality and endemic disease in the ancient city of Rome. *Ancient Society, 25*, 151-175.

Schneider, J. (2015). Null hypothesis significance tests. A mix-up of two different theories: the basis for widespread confusion and numerous misinterpretations. *Scientometrics, 102*(1), 411-432. doi: 10.1007/s11192-014-1251-5.

Seglen, P. O. (1992). The skewness of science. *Journal of the American Society for Information Science, 43*(9), 628-638.

Shaver, J. M. (2005). Testing for mediating variables in management research: Concerns, implications, and alternative strategies. *Journal of Management, 31*(3), 330-353. doi: 10.1177/0149206304272149.

Shockley, W., & Queisser, H. J. (1961). Detailed Balance Limit of Efficiency of P-N Junction Solar Cells. *Journal of Applied Physics, 32*(3), 510-&. doi: 10.1063/1.1736034.

Simon, H. A. (1956). Rational Choice and the Structure of the Environment. *Psychological Review, 63*(2), 129-138. doi: 10.1037/h0042769.

Simon, H. A. (1990). Invariants of Human-Behavior. *Annual Review of Psychology, 41*, 1-19. doi: DOI 10.1146/annurev.ps.41.020190.000245.

Smith, E. R. (2012). Editorial. *Journal of Personality and Social Psychology, 102*(1), 1-3.

Steinle, F. (2008). Explorieren - Entdecken - Testen. (9), 34-41.

Stephens, D. W., & Krebs, J. R. (1986). *Foraging theory*. Princeton, N.J.: Princeton University Press.

Tahamtan, I., & Bornmann, L. (2018). Core elements in the process of citing publications: Conceptual overview of the literature. *Journal of Informetrics, 12*(1), 203-216. doi: 10.1016/j.joi.2018.01.002.

Thaler, R. H., & Sunstein, C. R. (2008). *Nudge: improving decisions about health, wealth, and happiness*. New Haven, CT, USA: Yale University Press.

Thonon, F., Boulkedid, R., Delory, T., Rousseau, S., Saghatchian, M., van Harten, W., . . . Alberti, C. (2015). Measuring the Outcome of Biomedical Research: A Systematic Literature Review. *Plos One, 10*(4). doi: 10.1371/journal.pone.0122239.

Thor, A., Marx, W., Leydesdorff, L., & Bornmann, L. (2016). Introducing CitedReferencesExplorer (CRExplorer): A program for Reference Publication Year Spectroscopy with Cited References Standardization. *Journal of Informetrics, 10*(2), 503-515.

Tressoldi, P. E., Giofre, D., Sella, F., & Cumming, G. (2013). High impact = high statistical standards? not necessarily so. *Plos One, 8*(2). doi: 10.1371/journal.pone.0056180.

van Raan, A., van Leeuwen, T., & Visser, M. (2011). Severe language effect in university rankings: particularly Germany and France are wronged in citation-based rankings. *Scientometrics, 88*(2), 495-498. doi: 10.1007/s11192-011-0382-1.

van Raan, A. F. J. (2004). Sleeping Beauties in science. *Scientometrics, 59*(3), 467-472.

van Raan, A. F. J. (2008). Bibliometric statistical properties of the 100 largest European research universities: prevalent scaling rules in the science system. *Journal of the American Society for Information Science and Technology, 59*(3), 461-475. doi: 10.1002/asi.20761.

Vindolanda Tablets Online. (2018). Vindolanda Tablets Online (Tablet 154). Retrieved March 22, 2018, from http://vindolanda.csad.ox.ac.uk/

Vinkler, P. (2010). *The evaluation of research by scientometric indicators*. Oxford, UK: Chandos Publishing.

von Neumann, J., & Morgenstern, O. (1947). *Theory of games and economic behavior* (2. ed.). Princeton, NJ: Princeton University Press.

Wachter, R. (2016). How Measurement Fails Doctors and Teachers. Retrieved June 23, 2016, from http://www.nytimes.com/2016/01/17/opinion/sunday/how-measurement-fails-doctors-and-teachers.html?_r=1

Waltman, L. (2016). A review of the literature on citation impact indicators. *Journal of Informetrics, 10*(2), 365-391.

Waltman, L., & van Eck, N. J. (2016). The need for contextualized scientometric analysis: An opinion paper. In I. Ràfols, J. Molas-Gallart, E. Castro-Martínez & R. Woolley (Eds.),

  *Proceedings of the 21 ST International Conference on Science and Technology Indicator* (pp. 541-549). València, Spain: Universitat Politècnica de València.

Wildgaard, L., Schneider, J., & Larsen, B. (2014). A review of the characteristics of 108 author-level bibliometric indicators. *Scientometrics, 101*(1), 125-158. doi: 10.1007/s11192-014-1423-3.

Wilsdon, J., Allen, L., Belfiore, E., Campbell, P., Curry, S., Hill, S., . . . Johnson, B. (2015). *The Metric Tide: Report of the Independent Review of the Role of Metrics in Research Assessment and Management*. Bristol, UK: Higher Education Funding Council for England (HEFCE).

Ziliak, S. T., & McCloskey, D. N. (2004). Size matters: The standard error of regressions in the American Economic Review. *Econ Journal Watch, 1*(2), 331-U144.

Ziman, J. (2000). *Real science. What it is, and what it means*. Cambridge, UK: Cambridge University Press.